\def\ecrit{{E_{\rm crit}}}
\tolerance=10000
\catcode`@=11 

%
%
%
\catcode`@=11 
%
%
\font\seventeenrm=cmr17  
\font\fourteenrm=cmr12 scaled\magstep1  
\font\twelverm=cmr12  
\font\ninerm=cmr9

\font\sevenrm=cmr7
\font\sixrm=cmr6
\font\fiverm=cmr5
\font\seventeenbf=cmbx12 scaled\magstep2  
\font\fourteenbf=cmbx12 scaled\magstep1  
\font\twelvebf=cmbx12   
\font\ninebf=cmbx9

\font\sevenbf=cmbx7
\font\sixbf=cmbx6
\font\fivebf=cmbx5
\font\seventeeni=cmmi12 scaled\magstep2  
                                            \skewchar\seventeeni='177
\font\fourteeni=cmmi12 scaled\magstep1  
                                            \skewchar\fourteeni='177
\font\twelvei=cmmi12 
                                            \skewchar\twelvei='177
\font\ninei=cmmi9                           \skewchar\ninei='177
\font\eighti=cmmi8                          \skewchar\eighti='177
\font\seveni=cmmi7                          \skewchar\seveni='177
\font\sixi=cmmi6                            \skewchar\sixi='177
\font\fivei=cmmi5                           \skewchar\fivei='177
\font\seventeensy=cmsy10 scaled\magstep3    \skewchar\seventeensy='60
\font\fourteensy=cmsy10 scaled\magstep2     \skewchar\fourteensy='60
\font\twelvesy=cmsy10 scaled\magstep1       \skewchar\twelvesy='60
\font\ninesy=cmsy9                          \skewchar\ninesy='60
\font\eightsy=cmsy8                         \skewchar\eightsy='60
\font\sevensy=cmsy7                         \skewchar\sevensy='60
\font\sixsy=cmsy6                           \skewchar\sixsy='60
\font\fivesy=cmsy5                          \skewchar\fivesy='60
\font\seventeenex=cmex10 scaled\magstep3
\font\fourteenex=cmex10 scaled\magstep2
\font\twelveex=cmex10 scaled\magstep1
%
\font\seventeensl=cmsl12 scaled\magstep2  
\font\fourteensl=cmsl12 scaled\magstep1  
\font\twelvesl=cmsl12   
\font\ninesl=cmsl9

\font\seventeenit=cmti12 scaled\magstep2  
\font\fourteenit=cmti12 scaled\magstep1  
\font\twelveit=cmti12  
\font\tenit=cmti10
\font\nineit=cmti9

\font\seventeentt=cmtt12 scaled\magstep2  
\font\fourteentt=cmtt12 scaled\magstep1  
\font\twelvett=cmtt12  
\font\tentt=cmtt10
\font\ninett=cmtt9

%
\font\seventeencp=cmcsc10 scaled\magstep3
\font\fourteencp=cmcsc10 scaled\magstep2
\font\twelvecp=cmcsc10 scaled\magstep1
\font\tencp=cmcsc10
\newfam\cpfam
\newdimen\b@gheight		\b@gheight=12pt
\newcount\f@ntkey		\f@ntkey=0
\def\f@m{\afterassignment\samef@nt\f@ntkey=}
\def\samef@nt{\fam=\f@ntkey \the\textfont\f@ntkey\relax}
\def\rm{\f@m0 }
\def\mit{\f@m1 }         
\def\cal{\f@m2 }
\def\it{\f@m\itfam}
\def\sl{\f@m\slfam}
\def\bf{\f@m\bffam}
\def\tt{\f@m\ttfam}
\def\caps{\f@m\cpfam}

\def\seventeenpoint{\relax
    \textfont0=\seventeenrm          \scriptfont0=\fourteenrm
      \scriptscriptfont0=\twelverm
    \textfont1=\seventeeni           \scriptfont1=\fourteeni
      \scriptscriptfont1=\twelvei
    \textfont2=\seventeensy          \scriptfont2=\fourteensy
      \scriptscriptfont2=\twelvesy
    \textfont3=\seventeenex          \scriptfont3=\fourteenex
      \scriptscriptfont3=\twelveex
    \textfont\itfam=\seventeenit     \scriptfont\itfam=\fourteenit
    \textfont\slfam=\seventeensl     \scriptfont\slfam=\fourteensl
    \textfont\bffam=\seventeenbf     \scriptfont\bffam=\fourteenbf
      \scriptscriptfont\bffam=\twelvebf
    \textfont\ttfam=\seventeentt
    \textfont\cpfam=\seventeencp
    \samef@nt
    \b@gheight=17pt
    \setbox\strutbox=\hbox{\vrule height 0.85\b@gheight
				depth 0.35\b@gheight width\z@ }}

\def\fourteenpoint{\relax
    \textfont0=\fourteenrm          \scriptfont0=\tenrm
      \scriptscriptfont0=\sevenrm
    \textfont1=\fourteeni           \scriptfont1=\teni
      \scriptscriptfont1=\seveni
    \textfont2=\fourteensy          \scriptfont2=\tensy
      \scriptscriptfont2=\sevensy
    \textfont3=\fourteenex          \scriptfont3=\twelveex
      \scriptscriptfont3=\tenex
    \textfont\itfam=\fourteenit     \scriptfont\itfam=\tenit
    \textfont\slfam=\fourteensl     \scriptfont\slfam=\tensl
    \textfont\bffam=\fourteenbf     \scriptfont\bffam=\tenbf
      \scriptscriptfont\bffam=\sevenbf
    \textfont\ttfam=\fourteentt
    \textfont\cpfam=\fourteencp
    \samef@nt
    \b@gheight=14pt
    \setbox\strutbox=\hbox{\vrule height 0.85\b@gheight
				depth 0.35\b@gheight width\z@ }}

\def\twelvepoint{\relax
    \textfont0=\twelverm          \scriptfont0=\ninerm
      \scriptscriptfont0=\sixrm
    \textfont1=\twelvei           \scriptfont1=\ninei
      \scriptscriptfont1=\sixi
    \textfont2=\twelvesy           \scriptfont2=\ninesy
      \scriptscriptfont2=\sixsy
    \textfont3=\twelveex          \scriptfont3=\tenex
      \scriptscriptfont3=\tenex
    \textfont\itfam=\twelveit     \scriptfont\itfam=\nineit
    \textfont\slfam=\twelvesl     \scriptfont\slfam=\ninesl
    \textfont\bffam=\twelvebf     \scriptfont\bffam=\ninebf
      \scriptscriptfont\bffam=\sixbf
    \textfont\ttfam=\twelvett
    \textfont\cpfam=\twelvecp
    \samef@nt
    \b@gheight=12pt
    \setbox\strutbox=\hbox{\vrule height 0.85\b@gheight
				depth 0.35\b@gheight width\z@ }}
\def\tenpoint{\relax
    \textfont0=\tenrm          \scriptfont0=\sevenrm
      \scriptscriptfont0=\fiverm
    \textfont1=\teni           \scriptfont1=\seveni
      \scriptscriptfont1=\fivei
    \textfont2=\tensy          \scriptfont2=\sevensy
      \scriptscriptfont2=\fivesy
    \textfont3=\tenex          \scriptfont3=\tenex
      \scriptscriptfont3=\tenex
    \textfont\itfam=\tenit     \scriptfont\itfam=\seveni
    \textfont\slfam=\tensl     \scriptfont\slfam=\sevenrm
    \textfont\bffam=\tenbf     \scriptfont\bffam=\sevenbf
      \scriptscriptfont\bffam=\fivebf
    \textfont\ttfam=\tentt
    \textfont\cpfam=\tencp
    \samef@nt
    \b@gheight=10pt
    \setbox\strutbox=\hbox{\vrule height 0.85\b@gheight
				depth 0.35\b@gheight width\z@ }}
\def\ninepoint{\relax
    \textfont0=\ninerm          \scriptfont0=\sevenrm
      \scriptscriptfont0=\fiverm
    \textfont1=\ninei           \scriptfont1=\seveni
      \scriptscriptfont1=\fivei
    \textfont2=\ninesy          \scriptfont2=\sevensy
      \scriptscriptfont2=\fivesy
    \textfont3=\tenex          \scriptfont3=\tenex
      \scriptscriptfont3=\tenex
    \textfont\itfam=\nineit     \scriptfont\itfam=\seveni
    \textfont\slfam=\ninesl     \scriptfont\slfam=\sevenrm
    \textfont\bffam=\ninebf     \scriptfont\bffam=\sevenbf
      \scriptscriptfont\bffam=\fivebf
    \textfont\ttfam=\ninett
    \textfont\cpfam=\tencp
    \samef@nt
    \b@gheight=9pt
    \setbox\strutbox=\hbox{\vrule height 0.85\b@gheight
				depth 0.35\b@gheight width\z@ }}

\parindent=10pt
\parskip=0pt
\def\today{\ifcase\month\or
   January\or February\or March\or April\or May\or June\or
   July\or August\or September\or October\or November\or December\fi
   \space\number\day, \number\year}
\newdimen\pagewidth \newdimen\pageheight \newdimen\textheight
\newdimen\headheight \newdimen\footheight
\hsize=7.125truein
\hoffset=-0.25truein
\vsize=10.5truein
\voffset=-0.65truein
\baselineskip=11pt
\pagewidth=\hsize \pageheight=\vsize
\headheight=20pt
\textheight=708pt 
\footheight=16pt
\maxdepth=2pt

\font\largeheadfont=cmcsc10 
\headline={\hbox to\pagewidth{%
{\largeheadfont Princeton University \hskip105pt 
     March 9, 1987 \hfil DOE/ER/3072-41}}}
\nopagenumbers 
\footline={\hbox to\pagewidth{\hskip140pt%
Submitted to the 1987 Particle Accelerator Conference%
\hfil\folio}}
\def\evenpage{\hbox to\pagewidth{\folio\hfil}}
\def\oddpage{\hbox to\pagewidth{\hfil\folio}}
\def\onepageout#1{\shipout\vbox{
      \offinterlineskip 
      \vbox to\headheight{\the\headline \vskip3pt \hrule height1pt \vfill}
      \vbox to\textheight{#1} 
      \vbox to\footheight{\vfil 
              \ifnum\pageno=1\the\footline \else{\ifodd\pageno\oddpage
              \else\evenpage \fi}\fi}}
      \advancepageno}
\newbox\partialpage
\def\begindoublecolumns{\begingroup
       \output={\global\setbox\partialpage=\vbox{\unvbox255\bigskip}}\eject
       \output={\doublecolumnout} \hsize=3.4375truein \vsize=1416pt}
\def\enddoublecolumns{\output={\balancecolumns}\eject
       \endgroup \pagegoal=\vsize}
\def\doublecolumnout{\splittopskip=\topskip \splitmaxdepth=\maxdepth
       \dimen@=700pt \advance\dimen@ by-\ht\partialpage
       \setbox0=\vsplit255 to\dimen@ \setbox2=\vsplit255 to\dimen@
       \onepageout\pagesofar \unvbox255 \penalty\outputpenalty}
\def\pagesofar{\unvbox\partialpage
       \wd0=\hsize \wd2=\hsize \hbox to\pagewidth{\box0\hfil\box2}}
\def\balancecolumns{\setbox0=\vbox{\unvbox255} \dimen@=\ht0
       \advance\dimen@ by\topskip \advance\dimen@ by-\baselineskip
       \divide\dimen@ by2 \splittopskip=\topskip
       {\vbadness=10000 \loop \global\setbox3=\copy0
           \global\setbox1=\vsplit3 to\dimen@
           \ifdim\ht3>\dimen@ \global\advance\dimen@ by1pt \repeat}
       \setbox0=\vbox to\dimen@{\unvbox1} \setbox2=\vbox to\dimen@{\unvbox3}
       \pagesofar}
\centerline{\tenpoint\bf THE HAWKING-UNRUH TEMPERATURE}
\smallskip
\centerline{\bf AND QUANTUM FLUCTUATIONS IN PARTICLE ACCELERATORS}
\medskip
\centerline{K. T. McDonald}
\centerline{\sl Joseph Henry Laboratories, Princeton University,
Princeton, New Jersey 08544}
\medskip
\parindent=10pt
\begindoublecolumns
We wish to draw attention to a novel view of the effect of the quantum
fluctuations during the radiation of accelerated particles, particularly
those in storage rings.  This view is inspired by the remarkable insight
of Hawking$^1$ that the effect of the strong gravitational field of a black
hole on the quantum fluctuations of the surrounding space is to cause 
the black hole to radiate with a temperature
$$T = {\hbar g \over 2\pi ck},$$
where $g$ is the acceleration due to gravity at the surface of the black hole,
$c$ is the speed of light, and $k$ is Boltzmann's constant.  Shortly
thereafter Unruh$^2$ argued that an accelerated observer should become excited
by quantum fluctuations to a temperature
$$T = {\hbar a^\star \over 2\pi ck},$$
where $a^\star$ is the acceleration of the observer in its instantaneous rest
frame.  In a series of papers Bell and co-workers$^{3-5}$ have noted that
electron storage rings provide a demonstration of the utility of the 
Hawking-Unruh temperature, with emphasis on the question of the 
incomplete polarization of the electrons due to quantum fluctuations of
synchrotron radiation.

Here we expand slightly on the results of Bell {\it et al.}, and encourage the
reader to consult the literature for more detailed understanding.
\medskip
\centerline{\bf Applicability of the Idea}
\medskip
When an accelerated charge radiates, the discrete energy and momentum of the
radiated photons induce fluctuations on the motion of the charge.  The insight
of Unruh is that for uniform linear acceleration (in the absense of the 
fluctuations), the fluctuations would excite any internal degrees of freedom
of the charge to the temperature stated above.  His argument is very general
({\it i.e.}, thermodynamic) in that it does not depend on the details of
the accelerating force, nor of the nature of the accelerated particle.  The
idea of an effective temperature is strictly applicable only for uniform
linear acceleration, but should be approximately correct for other 
accelerations, such as that due to uniform circular motion.

A charged particle whose motion is confined by the focusing system of a 
particle accelerator exhibits transverse and longitudinal oscillations
about its ideal path.  These oscillations are excited by the quantum
fluctuations of the particle's radiation, and thus provide an excellent physical
example of the viewpoint of Unruh.

Further, the particles take on a thermal distribution of energies when
viewed in the average rest frame of a bunch, which transforms to the
observed energy spread in the laboratory.  While classical synchrotron
radiation would eventually polarize the spin-$1 \over 2$ particles completely,
the thermal fluctuations oppose this, reducing the maximum beam polarization.

It is suggestive to compare the excitation energy $U^\star = kT$, as would be
observed in the particle's rest frame, to the rest energy $mc^2$ when 
the acceleration is due to laboratory electromagnetic fields $E$ and $B$.  
Noting that $a^\star = eE^\star/m$ we find
$${U^\star \over mc^2} = {\hbar eE^\star \over 2\pi m^2c^3} = {\left[ 
E_\parallel + \gamma\left( E_\perp + \beta B_\perp \right)\right] \over 2\pi
\ecrit},$$
where the particle's laboratory momentum is $\gamma\beta mc$, and
$$\ecrit \equiv {m^2c^3 \over e\hbar}.$$
For an electron,
$$\ecrit = 1.3 \times 10^{16}\,{\rm volts/cm} =
4.4 \times 10^{13}\,{\rm gauss}.$$
($\ecrit$ is the field strength at which spontaneous pair production becomes
highly probable, {\it i.e.}, the field whose voltage drop across a
Compton wavelength is the particle's rest energy.)  We might expect that the
fluctuations become noticeable when $U^\star \sim 0.1$ eV, and hence comparable
to any other thermal effects in the system, such as the particle-source
temperature.

For linear accelerators $E_\parallel \sim 10^6$ volts/cm at best, so $U^\star
< 10^{-5}$ eV.  The effect of quantum fluctuations is of course negligible
because the radiation itself is of little importance in a linear accelerator.

For an electron storage ring such as LEP, $\gamma \sim 10^5$, and $B_\perp \sim
10^3$ gauss, so that $U^\star \sim 0.2$ eV.  For the SSC proton storage ring,
$\gamma \sim 2\times 10^4$, while $B_\perp \sim 6 \times 10^4$ gauss, so that
$U^\star \sim 2$ eV.  As is well known, in essentially all electron storage
rings, and in future proton rings, the effect of quantum fluctuations is
quite important.

The remaining discussion is restricted to beams in storage rings (= transverse
particle accelerators).
\medskip
\centerline{\bf Beam-Energy Spread}
\medskip
An immediate application of the excitation energy $U^\star$ is to the 
beam-energy spread.  In the average rest frame of a bunch of particles, the
distribution of energies is approximately thermal, with characteristic
kinetic energy $U^\star$, and momentum $p^\star = \sqrt{2mU^\star}$.  The
spread in laboratory energies is then given by
$$U_{\rm lab}  \approx \gamma (mc^2 + U^\star \pm \beta p^\star c) 
\approx U_0 \left( 1 \pm \gamma \sqrt{\lambda_C \over \pi\rho} \right),$$
where $U_0 = \gamma mc^2$ is the nominal beam energy,
$\rho = U_0/eB_\perp$ is the radius of curvature of the central
orbit, and $\lambda_C = \hbar/mc$ is the Compton wavelength. 
Writing this as
$$\left( {\delta U \over U_0}\right)^2 \approx {\gamma^2\lambda_C \over 
\pi\rho},$$
we obtain the standard result, as given by equation (5.48) of the review by
Sands.$^6$
\medskip
\centerline{\bf Beam Height}
\medskip
The quantum fluctuations of synchrotron radiation drive the oscillations of
particles about the bunch center, and set lower limits on the transverse and
longitudinal beam size.  If we associate a harmonic oscillator with each
component of the motion about the bunch center, then each oscillator will
be excited to amplitudes whose corresponding energy is $U^\star = kT^\star$.

For example, consider the vertical betatron oscillations which determine the 
beam height.  The frequency of these oscillations is $\omega = \nu_z \omega_0
= \nu_z c/R$, where $\nu_z$ is the vertical betatron number, and $R = L/2\pi$
is the mean radius of the storage ring.  In the average rest frame of a bunch
the oscillation frequency appears to be $\omega^\star = \gamma\omega$, and
the spring constant in this frame is given by $k^\star = m\omega^{\star 2} =
\gamma^2 m\omega^2$.  The typical amplitude of oscillation in this frame is
then
$${1 \over 2}k^\star z^{\star2} \approx U^\star = {\hbar a^\star \over 
2\pi c} = {\hbar\gamma^2 a \over 2\pi c} = {\hbar\gamma^2 c \over 2\pi\rho},$$
noting that in uniform circular motion the acceleration is transverse.
For the vertical oscillation
the lab frame amplitude $z$ is the same as $z^\star$.  Combining the above
we find
$$z^2 = {\lambda_C R^2 \over \pi\nu_z^2 \rho},$$
which reproduces the standard result, such as equation (5.107) of Sands.$^6$

An analogous argument is given in ref. 5 to derive the beam height in a weakly 
focused storage ring.
\medskip
\centerline{\bf Bunch Length and Beam Width}
\medskip
A similar analysis can be given for oscillations in the plane of the orbit.
However, radial and longitudinal excursions are also directly coupled to energy
excursions, which proves to be the stronger effect.  As the present method
finds the standard result for the beam-energy spread, the usual results for
bunch length and beam width follow at once.  [In ref. 6, use equations (5.64)
and (5.93) to yield expressions (5.65) and (5.95).]
\medskip
\centerline{\bf Beam Polarization}
\medskip
Sokolov and Ternov$^7$ predicted that quantum fluctuations in synchrotron
radiation limit the transverse polarization of the beam to 92\%.  In the absense
of quantum fluctuations the polarization should reach 100\% after long times.
Bell and Leinaas$^3$ realized that the thermal character
of the fluctuations provides an alternate view of the depolarizing mechanism.
In ref. 5 they provide a detailed justification that the
thermodynamic arguments are fully equivalent to the original QED calculation
of Sokolov and Ternov.  In the process they find that for circular motion
in a weakly focused ring (betatron), the effective temperature due to 
quantum fluctuations is
$$kT = {13 \over 96} \sqrt{3} {\hbar a^\star \over c},$$
which is about 1.5 times Unruh's result for linear acceleration.
\medskip
\centerline{\bf Radiation Spectrum}
\medskip
Because of the quantum fluctuations the motion of the particles departs from
the central orbit, and a classical calculation of the synchrotron-radiation
spectrum is incorrect in principle.  The deviations become significant only
when the characteristic energy of the radiation approaches the beam energy,
{\it i.e.}, when $\gamma B_\perp/\ecrit \sim 1$, and the prominent effect is
the cutoff at the high-energy end of the spectrum.

In the regime where the quantum corrections to the radiation spectrum are 
small the author has given an estimate of their size.$^8$  For this we
imagine the accelerated charge is surrounded (in its rest frame) by a
bath of photons with a Planck spectrum of temperature 
$kT = \hbar a^\star/2\pi c$.  The correction to the classical spectrum is
considered to arise from the Thomson scattering of these virtual photons off
the charged particle.  
In the lab frame the spectral correction is proportional to the Lorentz 
transform of the Planck
spectrum, whose peak photon energy is then $2\gamma kT = \hbar\gamma^3 c/
\pi\rho$, essentially the same as that of the classical spectrum.  
On integrating over energy, the
total rate of the correction term is the classical (Larmor) rate times
$${\alpha \over 60\pi} \left( {\gamma B_\perp \over \ecrit} \right)^2,$$
which is indeed very small at present storage rings.
\medskip
\centerline{\bf Acknowledgements}
\medskip
I would like to thank Ian Affleck and Heinrich Mitter for several discussions
on this topic.  This work was supported in part by the U.S. Department of Energy
under contract DOE-AC02-76ER-03072.
\medskip
\hbox to \hsize{\hss\vrule width2.5cm height1pt \hss}
\medskip
$^{1}$
S.W.~Hawking, 
``Black-Hole Explosions", 
Nature {\bf 248}, 30-31 (1974).

$^{2}$
W.G.~Unruh, 
``Notes on Black-Hole Evaporation",
Phys.\ Rev.\ D {\bf 14}, 870-892 (1976).

$^{3}$
J.S.~Bell and J.M.~Leinaas, 
``Electrons as Accelerated Thermometers",
Nucl.\ Phys.\ {\bf B212}, 131-150 (1983).

$^{4}$
J.S.~Bell, R.J.~Hughes and J.M.~Leinaas, 
``The Unruh Effect in Extended Thermometers", 
Z.\ Phys.\ C {\bf 28}, 75-80 (1985).

$^{5}$
J.S.~Bell and J.M.~Leinaas, 
``The Unruh Effect and Quantum Fluctuations of Electrons in Storage Rings", 
Nucl.\ Phys.\ {\bf B284}, 488-508 (1987).

$^{6}$
M.~Sands, 
``The Physics of Electron Storage Rings",
SLAC-121, (1970); also in Proc.\ 1969 Int.\ School of Physics, `Enrico
Fermi,' ed.\ by B.\ Touschek (Academic Press, 1971), p.~257.

$^{7}$
A.A.~Sokolov and I.M.~Ternov, 
``On Polarization and Spin Effects in the Theory of Synchrotron Radiation",
Sov.\ Phys.\ Dokl.\ {\bf 8}, 1203 (1964).

$^{8}$
K.T.~McDonald, 
``Fundamental Physics During Violent Acceleration",
in {\sl Laser Acceleration of Particles}, 
AIP Conf.\ Proc.\ {\bf 130} 23-54 (1985).

\enddoublecolumns
\output{\onepageout{\unvbox255}}
\vfill\eject\end